\documentclass[twocolumn,showpacs,preprintnumbers,amsmath,amssymb]{revtex4}
\usepackage{graphicx}
\begin{document}
\title{Tidal effects and the Proximity decay of nuclei}
\author{A.B. McIntosh}
\author{S. Hudan}
\author{C.J. Metelko}
\author{R.T. de Souza} 
\affiliation{
Department of Chemistry and Indiana University Cyclotron Facility \\ 
Indiana University, Bloomington, IN 47405} 

\author{R.J. Charity}
\author{L.G. Sobotka}
\affiliation{
Department of Chemistry, Washington University, St. Louis, MO 63130}

\author{W.G. Lynch}
\author{M.B. Tsang} 
\affiliation{
National Superconducting Cyclotron Laboratory and Department of
Physics and Astronomy \\ 
Michigan State University, East Lansing, MI 48824}

\date{\today}

\begin{abstract}
We examine the decay of the 3.03 MeV state of $^8$Be evaporated from an 
excited 
projectile-like fragment following a peripheral heavy-ion collision. 
The relative energy of the daughter $\alpha$ particles exhibits a 
dependence on the decay angle of the $^8$Be$^*$, indicative of a tidal effect. 
Comparison of the
measured tidal effect with a purely Coulomb model suggests the influence of 
a measurable nuclear proximity interaction.

\end{abstract}
\maketitle

Aggregation of clusters in a dilute medium is a process that impacts
a wide range of physical phenomena from the formation of 
galactic structure\cite{***} 
to formation of 
Van der Waals clusters in a low density gas\cite{***}. 
This aggregation can involve the delicate interplay of elementary forces
that results in a frustrated system. One such example is the formation of pasta
nuclei in the crust of a neutron star\cite{Ravenhall83,Horowitz04}.
For smaller systems, the phenomenon of alpha clustering
is important both in low density nuclear matter\cite{Ropke98},
as well as in light nuclei\cite{Tohsaki01}.
In the case of heavy nuclei,
cluster aggregation is manifested in
the spontaneous phenomenon of clusters, from the common
process of $\alpha$ decay to the emission of more exotic 
clusters such as $^{14}$C \cite{Rose84}. 
The reduction of density near the nuclear surface, 
allows formation of $\alpha$ particles, or other 
clusters, and their emission from 
either ground-state or modestly excited nuclei. Cluster emission,
thus primarily probes the surface properties of the emitting 
nucleus\cite{Sobotka06}.
Our present understanding of cluster emission is largely based upon 
the yields, 
kinetic energy spectra, and angular distributions of emitted clusters - 
all of which are well 
described within a statistical transition-state formalism\cite{Moretto75}.
In this Letter we present for the first time 
evidence for the
modification of cluster emission by interaction
with the nuclear surface. 
We probe the interaction of the nuclear surface with the emitted
cluster by using resonance spectroscopy to examine the emission of 
$^8$Be$^*$ and specifically explore how its decay is impacted by the 
tidal effect\cite{Charity01}. 

Charged-particles produced in the reaction $^{114}$Cd+$^{92}$Mo at E/A=50 MeV 
were detected in an 
exclusive 4$\pi$ setup. 
To focus on evaporated fragments, we selected
peripheral collisions through detection of 
forward-moving projectile-like fragments (PLFs) with 10$\le$Z$\le$48,
in the angular range 
2.1$^{\circ}$$\leq$$\theta^{lab}$$\leq$4.2$^{\circ}$
with $\Delta$$\theta^{lab}$$\approx$0.13$^\circ$\cite{Hudan05}.  
This PLF is the decay residue of the excited primary projectile-like 
fragment (PLF$^*$) formed by the collision.
Light-charged-particles and
fragments with Z$\le$9 were
isotopically identified\cite{Hudan05} in the angular range 
7$^{\circ}$$\leq$$\theta^{lab}$$\leq$58$^{\circ}$ with the 
silicon-strip array LASSA \cite{Davin01,Wagner01}. 
Each of the nine 
telescopes in this array consisted of a stack of three elements, two 
5cm x 5cm  silicon strip detectors (Si(IP)) backed by a
2 x 2 arrangement of
CsI(Tl) crystals each with photo-diode readout. 
Each telescope 
was segmented into 16x16 
orthogonal strips, resulting in good angular resolution 
($\Delta$$\theta^{lab}$$\approx$0.43$^\circ$). 
The 3 x 3 arrangement of the LASSA telescopes was centered 
at a polar angle 
$\theta^{lab}$=32$^\circ$ with respect to the beam axis.
The identification threshold of LASSA is 2 MeV/A for 
$\alpha$ particles. Light-charged particles emitted at other angles were
detected by the Miniball/Miniwall 4$\pi$ phoswich array.
For the analysis that follows 
events were selected with
15$\le$Z$_{PLF}$$\le$46, 
V$_{PLF}$$\ge$8.0cm/ns and the multiplicity of particles in LASSA, 
N$_{LASSA}$=2,3.
Using the 
measured emitted particles and the assumption of isotropy, the Z, A, 
and velocity of the PLF$^*$ was calculated\cite{Yanez03}. 
Associated with these events is a PLF$^*$ with a most probable atomic 
number of $\approx$30.

\begin{figure}
\vspace*{3.0in} 
\includegraphics{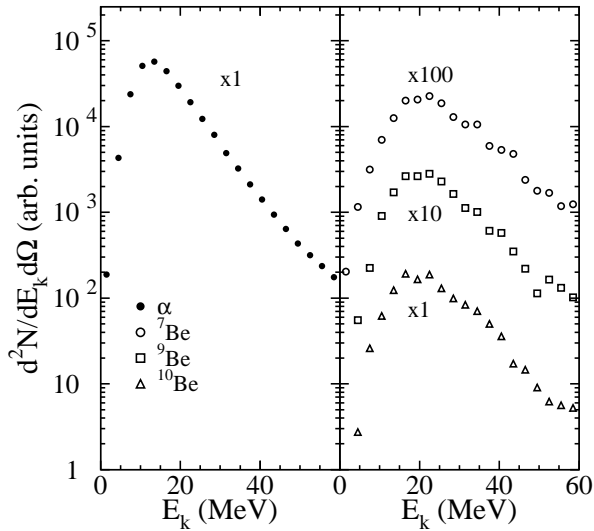}
\caption[]
{
Kinetic energy spectra of $\alpha$ particles and 
$^{7,9,10}$Be fragments in the PLF$^*$ frame observed with 
$\theta$$_{PLF*}$$\le$80$^\circ$.
} 
\label{fig:figure1}
\end{figure}

The kinetic energy spectra of $\alpha$ particles and Be nuclei emitted in the 
angular range
$\theta$$_{PLF*}$$\le$80$^\circ$
are presented in Fig.~\ref{fig:figure1}. 
These
spectra are reasonably well described by a Maxwell-Boltzmann distribution 
supporting the picture of a largely evaporative process\cite{Yanez03}.
In comparison to the $\alpha$ particle spectra, the spectra of Be fragments
exhibit more pronounced tails indicative of a higher initial 
temperature\cite{Yanez03}. The average kinetic energy of $^7$Be is larger
than that of $^{9,10}$Be in agreement with previous work suggestive of 
sequential decay of excited primary fragments
as they propagate away from the emitting nucleus\cite{Hudan05}. 
For the events presented, damping in the collision 
process produces a PLF$^*$ with a most probable excitation energy 
of 3 MeV/A and a maximum excitation of 4 MeV/A.
The deduced excitation is consistent
with the ``temperature'' associated with the $\alpha$ particle kinetic energy 
spectra\cite{Yanez03}.  
The de-excitation cascade of the PLF$^*$ involves the emission of 
not only nucleons and
ground state fragments, but also excited fragments, some of which are 
unstable against particle emission.

\begin{figure}[t]
\vspace*{3.0in} 
\includegraphics{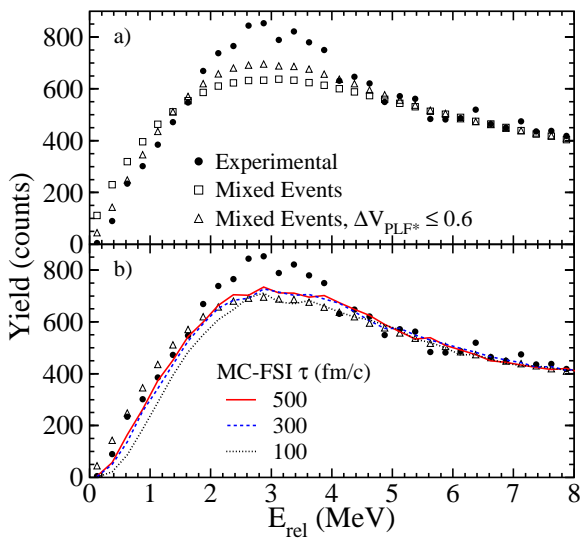}
\caption[]
{Panel a: Relative kinetic energy distributions for $\alpha$-$\alpha$ pairs. 
Panel b: Comparison of 
the results of a non-resonant final-state interaction Monte Carlo 
calculation with the mixed-event background.} 
\label{fig:figure2}
\end{figure}

The decay of short-lived resonant states such as $^8$Be$^*$ can be examined 
by constructing the relative energy spectrum of the daughter products. Shown
in Fig.~\ref{fig:figure2}a is the relative energy spectrum between two
$\alpha$ particles for events in which either two or three $\alpha$ particles
were detected in LASSA with $\theta_{\alpha,PLF*}$$\le$100$^\circ$. 
Solid symbols depict the experimental data which
exhibits a peak at E$_{rel}$$\approx$3 MeV. The 
overall shape of the E$_{rel}$ spectrum is affected at low
E$_{rel}$ ($\le$ 2 MeV) by the finite angular acceptance of 
the LASSA CsI(Tl) detectors,
as two particles entering the same CsI(Tl) crystal are not 
resolved. The $^8$Be ground-state is consequently not observed in this 
kinematic regime.
For larger E$_{rel}$ ($\ge$5 MeV), the decrease in yield is impacted by 
the geometric acceptance of LASSA. 
The observed relative energy distribution 
can be understood as
having two primary components: resonant decay of $^8$Be$^*$ and non-resonant 
$\alpha$ emission.  
In order to 
assess the non-resonant contribution, 
which acts as a  ``background'' for the resonant decay,
we performed a 
mixed-event analysis. Two $\alpha$ particles were selected from 
different events and their relative energy was calculated. The resulting 
relative energy spectrum was normalized in the interval 
14$\le$E$_{rel}$$\le$50 MeV (where the resonant decay contribution 
is expected to be insignificant) and is displayed as open 
squares in Fig.~\ref{fig:figure2}a. While the mixed-event background has the 
general shape of the observed relative energy spectrum, there is an 
excess yield in the observed data centered at $\approx$3 MeV. This excess 
originates from the decay of the first excited state of $^8$Be 
at 3.03 MeV with an intrinsic width of 1.5 MeV. 
Over-prediction of the yield for E$_{rel}$$\le$1.5 MeV is
understandable as the mixed-event pairs do not experience Coulomb 
repulsion. One of the 
drawbacks of the mixed-event analysis is that the events sampled span a 
broad distribution of PLF$^*$ velocities. To select the conditions that best describe 
the non-resonant $\alpha$ emission, we restricted the difference in 
PLF$^*$ velocity, $\Delta$V$_{PLF*}$, for the events in the mixed-event 
background. The mixed-event background corresponding to 
$\Delta$V$_{PLF*}$$\le$0.6 cm/ns is shown as open triangles in 
Fig.~\ref{fig:figure2}a. This background provides a better description of 
the data for E$_{rel}$$\le$1.5 MeV while it exhibits a larger yield
at E$_{rel}$$\approx$3 MeV. To determine which value of $\Delta$V$_{PLF*}$ 
provided the best reproduction of the non-resonant decay, we calculated the
$\chi$$^2$ 
between the mixed
events and the experimental data in the interval 6$\le$E$_{rel}$$\le$50 MeV. 
The minimum
$\chi$$^2$/$\nu$=1-1.1 corresponds to
$\Delta$V$_{PLF*}$=
0.4-0.6cm/ns. 

     In order to better understand the ``background'', we have 
modeled the non-resonant final-state interaction with a Monte Carlo Coulomb 
trajectory calculation. In this model, which we refer to as MC-FSI, 
two $\alpha$ particles are 
isotropically emitted in sequential fashion 
from the surface of an excited nucleus following Lambert emission,
while accounting for all recoil and Coulomb interactions. 
The Z$_{PLF*}$, A$_{PLF*}$, 
V$_{PLF*}$, and $\theta$$_{PLF*}$ are taken from the 
experimental data while the time between emissions is taken to be exponential 
with a mean time $\tau$. 
Following Coulomb propagation 
all particles were filtered for the experimental acceptance, 
angular resolution, and thresholds. Calculations with
$\tau$=100, 300, and 500 fm/c are shown by the lines 
in Fig.~\ref{fig:figure2}b along
with the experimental data and mixed-event background. With 
decreasing mean emission time, 
suppression of yield for low values of E$_{rel}$ is observed.
While the calculation with $\tau$=100 fm/c is
clearly inconsistent with the 
experimental data, $\tau$$\ge$300 fm/c provides a reasonable 
description of the mixed-event background for E$_{rel}$$\ge$2 MeV. 
A deduced mean emission time of this magnitude associated with an 
excitation energy of 3 MeV/A is consistent with previous work\cite{Beaulieu00}.
For smaller values of E$_{rel}$, mutual Coulomb repulsion of the 
two $\alpha$ particles results in a suppression of yield at small E$_{rel}$
as compared to the mixed-event data. 
The comparison of the mixed-event background with the
Monte Carlo final state calculation indicates that for all but the smallest
values of E$_{rel}$ the mixed-event background provides a reasonable 
description of the non-resonant contribution.

\begin{figure}
\vspace*{3.5in} 
\includegraphics{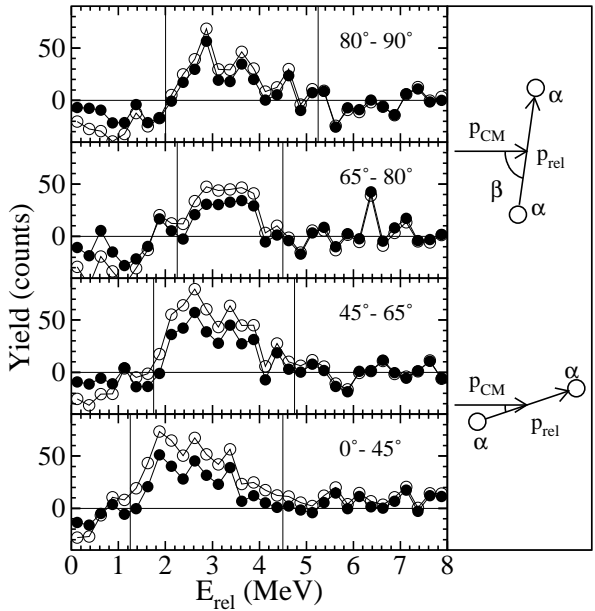}
\caption[]
{Difference relative energy spectra of observed $\alpha$ pairs for different
decay angles, $\beta$, relative
to the mixed-event background.} 
\label{fig:figure3}
\end{figure}

The first excited state of $^8$Be has an intrinsic width of 
1.5 MeV\cite{Tilley04} 
corresponding to a
mean lifetime of $\approx$131 fm/c. For such a short lifetime the likelihood
that 
this state decays in the vicinity of the emitting nucleus is significant. 
Consequently, as the $^8$Be$^*$ decays into two $\alpha$ particles, 
its increasing quadrupole moment interacts with the
gradient of the Coulomb field. 
For $\alpha$ pairs which decay orthogonal to the
emission direction, the field gradient provided by the emitting nucleus 
acts to 
increase their relative energy. In contrast, $\alpha$ pairs which decay 
along the emission direction experience a reduced relative energy due to the
larger acceleration of the nearer $\alpha$ particle\cite{Charity01}.
In order to examine this Coulomb tidal effect, we constructed the difference 
relative energy spectra of the observed 
$\alpha$ pairs relative
to the mixed-event background. These spectra were constructed for
different decay angles, $\beta$,  calculated as the angle between the
relative velocity of the $\alpha$-$\alpha$ pair
and the center-of-mass momentum of the pair. 
Depicted as the solid symbols in Fig.~\ref{fig:figure3} is the case using the
mixed-event background with $\Delta$V$_{PLF*}$$\le$0.6 cm/ns. 
For reference, the difference spectra with no restriction on
$\Delta$V$_{PLF*}$ are also shown as open symbols. Clearly the
$\Delta$V$_{PLF*}$ restriction does not introduce a major change in the 
difference relative energy spectra.
The prominent feature of the
difference spectra is the peak at $\approx$3 MeV which is evident 
for all $\beta$. 
The negative dip for E$_{rel}$$<$$\approx$1.5 MeV in the difference spectra 
is due to the over-prediction of the yield by the 
mixed-event background. 
Moreover, it is clearly evident that the peak in the difference spectrum for 
0$\le$$\beta$$\le$45$^\circ$ is shifted to lower values of E$_{rel}$ 
as compared to larger values of $\beta$. To quantitatively extract the
dependence of $<$E$_{rel}$$>$ on $\beta$ 
for the 3.03 MeV state,
we integrated the difference spectra
over the region indicated by the vertical lines. These integration limits 
were selected by considering the intrinsic width of the 3.03 MeV 
state and the 
shape of the measured spectra. The extracted $<$E$_{rel}$$>$ is
 largely insensitive to
any reasonable choice of integration limits.  
Using these difference spectra
and accounting for efficiency, we also 
extracted the total yield for the 3.03 MeV 
state. 
Based upon the yield of measured $^{7,9}$Be isotopes we estimate the yield
for the 3.03 MeV state. The extracted yield agrees 
with this expectation to within 15\%. 

\begin{figure}
\vspace*{3.0in} 
\includegraphics{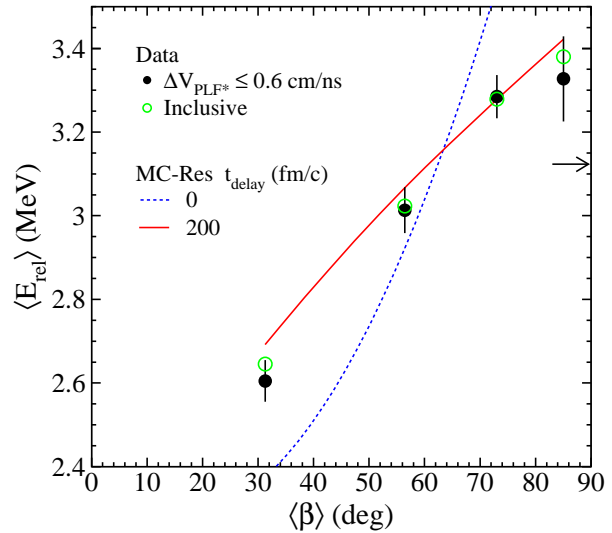}
\caption[]
{Dependence of $<$E$_{rel}$$>$, associated with 
the two $\alpha$ particles arising from the decay of first excited state of 
$^8$Be, on the average decay angle, $<$$\beta$$>$. Lines represent the 
results of Monte Carlo calculations for resonance decay.} 
\label{fig:figure4}
\end{figure}

The dependence of the average relative energy, $<$E$_{rel}$$>$, on 
the decay angle $\beta$ is shown in Fig.~\ref{fig:figure4} as solid symbols. 
A clear manifestation of the tidal effect is observed as 
$<$E$_{rel}$$>$ increases with increasing $\beta$. 
The magnitude of the 
observed change in $<$E$_{rel}$$>$
is $\approx$0.7 MeV, a relative change of $\approx$25$\%$. 
For reference, the arrow indicates the 
decay energy corresponding to the intrinsic energy of the state 
and the Q-value of the decay. The error bars shown, while dominated by the
measurement statistics, include the uncertainty associated with the 
integration limits. To demonstrate the
sensitivity of $<$E$_{rel}$$>$ to different mixed-event backgrounds 
we also show as open symbols 
the dependence of $<$E$_{rel}$$>$ on $\beta$ for no
restriction on $\Delta$V$_{PLF*}$. 
The same overall trend of 
$<$E$_{rel}$$>$ on $\beta$ is observed for the different mixed-event 
conditions.

To understand this observed trend quantitatively, we simulated the 
decay of a $^8$Be$^*$ in a simple Monte Carlo model, MC-Res. In this model the 
$^8$Be$^*$ is emitted isotropically from the surface of the PLF$^*$. 
The properties of the
PLF$^*$ were sampled in the same manner as in the MC-FSI calculations.
The lifetime
of the emitted $^8$Be$^*$ was chosen consistent with an exponential
probability distribution
P(t)=exp(-t/$\tau$), reflecting first order kinetics. The 
mean lifetime, $\tau$, was taken to be 131 fm/c determined by the 
intrinsic state width. 
The initial kinetic energy of the $^8$Be$^*$ is taken to be exponential with
a slope parameter of 7.5 MeV, consistent with the experimental 
data shown in Fig.~\ref{fig:figure1}. 
The $^8$Be$^*$ 
propagates in the
Coulomb field of the emitting nucleus until it decays. 
At the moment of decay the $^8$Be$^*$ 
is replaced with two $\alpha$ particles with an inter-$\alpha$ separation 
distance (scission configuration) 
given by R$_{\alpha - \alpha}$= 5.81 fm in accordance with 
systematics\cite{Sobotka83}. Selecting a smaller quadrupole moment, namely
R$_{\alpha - \alpha}$= 4 fm, makes a negligible difference in the 
final results. 
The decay angle, $\beta$, of the
two $\alpha$ system with respect to the emission direction is taken 
to be isotropic namely 
the effect of the Coulomb field in 
orienting the decaying $^8$Be$^*$ 
is neglected. Following decay, the two $\alpha$ 
particles are propagated along trajectories corresponding to both the 
Coulomb repulsion from the PLF$^*$ and their mutual Coulomb repulsion. 
Energy and momentum are conserved at all stages of the calculation. Particles
are subsequently filtered by the detector acceptance, 
angular resolution, and thresholds.

The results of this schematic calculation are shown as the dashed line in 
Fig.~\ref{fig:figure4}. The angular dependence of $<$E$_{rel}$$>$, namely
the magnitude of the tidal effect, is clearly overestimated by the 
purely Coulomb calculation. 
One possible reason for this difference 
is the interaction of the emitted $^8$Be$^*$ with the nuclear surface. 
This interaction is
particularly important as the most probable initial velocity of the 
$^8$Be$^*$ is 
close to zero, and consequently the $^8$Be$^*$ 
spends a significant time within the range of the proximity interaction. 
This is in marked contrast to the tidal decay 
following projectile breakup reactions\cite{charity95}.
The attractive surface interaction
can impact the decay of the evaporated $^8$Be$^*$ in two ways. It can 
stabilizes the 
$^8$Be$^*$ while it is in the vicinity of the emitting nucleus. 
In addition, the attractive potential may produce a {\em nuclear} tidal effect 
with the {\em opposite 
sign} as compared to the Coulomb tidal effect. 
Both the nuclear tidal effect and the increased
stability of the $^8$Be$^*$ will result in a decreased dependence of 
$<$E$_{rel}$$>$ on $\beta$.
To assess whether nuclear attraction could be responsible for the 
difference between the observed tidal effect and the Coulomb calculation,
we have introduced a delay into
the decay time distribution such that P(t)=exp(-(t-t$_{delay}$)/$\tau$) for  
t$>$t$_{delay}$. 
This artificial delay mimics
the increased stability of the $^8$Be$^*$ due to the 
nuclear interaction.
With increasing t$_{delay}$, the dependence of $<$E$_{rel}$$>$ on
$\beta$ decreases as evident in Fig.~\ref{fig:figure4}.
For t$_{delay}$=200 fm/c, represented by the solid line, 
one observes reasonable agreement with the 
observed tidal effect. Without any delay the lifetime of 131 fm/c 
corresponds to a 
decay distance of $\approx$17 fm. 
For t$_{delay}$=200 fm/c,
the $^8$Be$^*$ 
travels on average an additional 4 fm before decaying. From this we conclude
that stabilization of the $^8$Be$^*$ by the emitting nucleus for a distance
of $\approx$4 fm is sufficient to reproduce the magnitude of the 
observed tidal effect. A more realistic model of the nuclear interaction
is necessary in order to extract more quantitative results.

In summary, we have measured the dependence of $<$E$_{rel}$$>$ on 
decay angle for 
the first excited state of $^8$Be. This trend can be understood as 
a tidal effect in which 
the decaying cluster interacts with the gradient of an 
external field provided by the emitting nucleus. Calculations with
a Coulomb model over-predict the observed tidal effect. This over-prediction 
suggests that 
the emitting nucleus influences the cluster decay through the 
nuclear interaction. The attractive nuclear potential of the 
emitting nucleus acts to both 
stabilize the excited cluster and induces a nuclear tidal effect which 
opposes the Coulomb tidal effect.

\begin{acknowledgments}

	We would like to acknowledge the 
valuable assistance of the staff at MSU-NSCL for
providing the high quality beams which made this experiment possible. 
This work was supported by the
U.S. Department of Energy under DE-FG02-88ER40404 (IU), 
DE-FG02-87ER-40316 (WU) and the
National Science Foundation under Grant No. PHY-95-28844 (MSU).\par
\end{acknowledgments}

\bibliography{tidal1.bib} 

\end{document}